\documentclass[pra,preprint,showpacs,preprintnumbers,amsmath,amssymb]{revtex4}

\usepackage{graphicx}
\usepackage{epsf}
\usepackage{color}

\begin{document}
\title{The pinning quantum phase transition in a Tonks Girardeau gas: diagnostics by ground state fidelity and the Loschmidt echo}

\author{K.~Lelas,$^{1}$ T.~\v Seva,$^{2}$ H.~Buljan,$^{2}$ and J.~Goold$^{3,4}$}

\affiliation{$^1$Faculty of Electrical Engineering Mechanical
Engineering and Naval Architecture, University of Split, Rudjera
Bo\v skovi\' ca BB, 21000 Split, Croatia,} 
\affiliation{$^2$Department of Physics,
University of Zagreb, Bijenicka c. 32, 10000 Zagreb, Croatia,}
\affiliation{$^3$Clarendon Laboratory, University of Oxford, United Kingdom,}
\affiliation{$^4$Physics Department, University College Cork, Cork, Ireland}

\date{\today}

\begin{abstract}
We study the pinning quantum phase transition in a Tonks-Girardeau
gas, both in equilibrium and out-of-equilibrium, using the ground
state fidelity and the Loschmidt echo as diagnostic tools. The
ground state fidelity (GSF) will have a dramatic decrease when the
atomic density approaches the commensurate density of one particle
per lattice well. This decrease is a signature of the pinning
transition from the Tonks to the Mott insulating phase. We study the
applicability of the fidelity for diagnosing the pinning transition
in experimentally realistic scenarios. Our results are in excellent
agreement with recent experimental work. In addition, we explore the
out of equilibrium dynamics of the gas following a sudden quench
with a lattice potential. We find all properties of the ground
state fidelity are reflected in the Loschmidt echo dynamics i.e., in
the non equilibrium dynamics of the Tonks-Girardeau gas initiated by
a sudden quench of the lattice potential.




\end{abstract}

 \maketitle

\section{Introduction}

In the past two decades, ultra-cold atomic systems have emerged as
ideal playgrounds for the controlled simulation and manipulation of
textbook models from many-body physics \cite{Bloch:08}. Using the
full armory of developed tools the parameters of the underlying
Hamiltonian can be tuned with an unprecedented precision allowing
for the exploration of phase diagrams synonymous with condensed
matter physics. In addition, the high degree of isolation,
tunability and long coherent time scales associated with ensembles
of ultra-cold atoms allow for excellent time resolution of quantum
dynamics \cite{Cazalilla}.

For a long time studies of integrable systems were considered a
purely academic pursuit, but by now can be created in the laboratory
with ensembles of cold atoms. By applying the appropriate lasers to
Bose-Einstein condensates, one dimensional arrays of atoms may be
formed \cite{OneD}. In the limit of strong interactions these arrays
were observed to be in a fermionised state known as the
Tonks-Girardeau gas \cite{TG2004,Kinoshita2006}, a prototypical
integrable system.

In this work we will focus on the Tonks-Girardeau gas in a
particularly interesting configuration which admits critical point.
If a weak periodic potential is applied along the axial direction of
a one-dimensional ultra-cold quantum gas it is possible to generate
an atomic simulation of the Sine-Gordon model \cite{buchler}. When
the interactions between the particles in the gas are sufficiently
repulsive and the lattice is commensurate with the particle density
(one particle per lattice well) this model has a quantum phase
transition (at $T\approx0K$) where atoms become 'pinned' to the Mott
insulator state. In contrast to the well known superfluid-Mott
insulator transition, pinning to the Mott phase occurs for an
infinitesimally weak lattice potential \cite{buchler}. A spectacular
recent experiment demonstrated this transition for ensembles of one
dimensional ultra-cold gases \cite{Nagerl:10}.

In general, a quantum many-body system which undergoes a quantum
phase transition may be written as
$$\hat{H}(\lambda)=\hat{H_0}+\lambda\hat{H'},$$
where $\lambda$ and $\hat{H'}$ are the driving parameter and the
Hamiltonian driving the quantum phase transition (QPT) respectively.
A feature of a phase transition is that if the parameter $\lambda$
is varied across the critical point, the energy spectrum undergoes a
dramatic change i.e., the ground states of $\hat{H}(\lambda)$ and
$\hat{H}(\lambda+\delta\lambda)$ will significantly differ. As a
consequence the overlap of the ground states is expected to be
sensitive to a QPT \cite{Zanardi:06}. According to \cite{buchler},
the Tonks-Girardeau gas has a pinning quantum phase transition at
$\lambda=0$ when the driving Hamiltonian ($\hat{H'}$) includes an
optical lattice $V_l(x)=V_l\sin^2(kx)$ commensurate with atomic
density where the amplitude of the lattice is the parameter driving
QPT i.e., $V_l=\delta\lambda$. In this paper we shall denote
$\hat{H}(\lambda=0)$ as the Hamiltonian of TG gas in a trapping
potential $V_0(x)$  and $\hat{H}(\lambda+\delta\lambda=V_l)$ as the
Hamiltonian of TG gas in $V_0(x)+V_l(x)$ potential. We denote the
ground state of $\hat{H}(0)$ as $|\Psi_0\rangle$ and ground state of
$\hat{H}(V_l)$ as $|\Phi_0\rangle$. We expect that the overlap of
ground states $\langle\Psi_0|\Phi_0\rangle$ will be sensitive even
to infinitesimally weak optical lattice if lattice periodicity is
commensurate with atomic density \cite{buchler,Zanardi:06}. In
quantum information theory, the square modulus of the overlap is
known as fidelity \cite{quantinfo} and is a central concept in state
characterisation. The ground state fidelity is defined as
$$F=|\langle\Psi_0|\Phi_0\rangle|^2.$$


In this work we use this fidelity to study pinning quantum phase
transition in the Tonks-Girardeau gas. We find, as expected, that
GSF decreases with the increase of the lattice amplitude and size of
the system. We emphasize that in the thermodynamic limit the GSF can
unequivocally determine the pinning quantum phase transition for an
infinitesimally small lattice amplitude. Nevertheless, the auxiliary
trapping potential and finite size effects are important for
experimentally relevant numbers of particles. We find that the GSF
is in agreement with recent experiments on the pinning quantum phase
transition (QPT) for a Luttinger liquid of strongly interacting
bosons \cite{Nagerl:10}.
All of the observed properties of ground state fidelity are also
reflected in the dynamical evolution of the system i.e in survival
probability or the Loschmidt echo
\cite{Peres,Rodolfo,Jacquod,Cerruti,Prosen} (for a review see e.g.
\cite{Gorin}). The average value of the Loschmidt echo decreases for lower
value of ground state fidelity; that is a general observation.
Details of Loschmidt echo dynamics, such as the dominant frequency of
revivals, depend on the particular trapping potential. We find that for the
TG gas in an infinitely deep box potential oscillations of the
Loschmidt echo are large and occur with smaller frequency in the critical
region than in rest of parameter space. In the harmonic oscillator
potential, the frequency of the Loschmidt echo revivals is constant until
we reach a critical number of particles $N_{pinn}$, and after
$N_{pinn}$ the oscillations become irregular.

\section{The pinning transition in a Tonks-Girardeau gas}
\label{sec:quench}

Consider a gas of bosons confined in a tight waveguide at
$T\approx0K$ temperature with tight transverse trapping frequencies
such that $\omega_{\perp}\gg {\mu}/{\hbar}$, where $\mu$ is the
chemical potential. In this regime we may describe the many-body
system by an effective one dimensional Hamiltonian,
\begin{equation}
\label{eq:hamiltonian} \hat{H_0}=\int dx
\hat{\Psi}^{\dagger}(x)[\frac{-\hbar^{2}}{2m}\frac{d^2}{dx^2}+V_0(x)]\hat{\Psi}(x)+\frac{g}{2}\int
dx \hat{\Psi}^{\dagger}(x) \hat{\Psi}^{\dagger}(x) \hat{\Psi}(x)
\hat{\Psi}(x),
\end{equation}
where $V_0(x)$ is an arbitrary one dimensional longitudinal external
potential and $g$ describes the strength of a short ranged
interaction. In such one dimensional systems it is typical to
introduce the following dimensionless parameter,
$\gamma=mg/(\hbar^{2}\rho)$, which is the ratio of the kinetic
energy to the interaction energy ($\rho$ is the linear density). In
the spatially uniform case the spectrum is gapless for all $\gamma$
and described by a Luttinger liquid of bosons. Let us assume a one
dimensional optical lattice $V_{l}(x)=V_l\sin^{2}(kx)$ is applied
along the longitudinal direction of the waveguide in addition to
already existing trapping potential $V_0(x)$, in this case $V_l$ is
the strength of the applied lattice and we introduce wave vector
$k=2\pi/\lambda$. When interactions are weak, $\gamma\ll 1$, and the
lattice strength is much larger than the recoil energy $V_l\gg
E_{R}=(\hbar k)^2/(2m)$, Eq.~\eqref{eq:hamiltonian} may be mapped on
to the Bose-Hubbard model in the tight binding approximation
\cite{Bloch:08}. In this model, there is a phase transition as one
changes the ratio of tunneling to atom-atom interactions, between a
superfluid state where the atoms are free to tunnel between the
wells coherently and a Mott state with an excitation gap and fixed
number of particles per lattice site.

Interestingly, in the opposite case when the strength of the applied
lattice is much smaller than the recoil energy $V_l\ll E_{R}$, the
Bose-Hubbard model is not applicable as the bosons now occupy several
vibrational states in each well. In this case it was shown by
B\"uchler \textit{et al} that the system maybe mapped to the famous
Sine-Gordon model \cite{buchler}, an effective low energy theory has
been extensively studied in the literature as a rare example of an
exactly solvable quantum field theory. B\"uchler \textit{et al}
showed that when the gas is in strongly interacting Tonks Girardeau
limit, $\gamma\gg1$, and the lattice is commensurate with the
density then the system will be 'pinned' to the Mott insulator state
for an arbitrary weak lattice \cite{buchler}.

\section{The Fermi-Bose mapping theorem and ground state fidelity}
\label{sec:FB}

The pinning phase transition is quite straightforward to understand in the
Tonks-Girardeau limit of strong repulsive interactions,
$g\rightarrow\infty$, in which this work will focus on. Physically,
one may understand the pinning phase transition in this limit as the
competition between the average inter-particle distance due to the
strong interactions and the period of the potential. In this limit
the hard core interactions play the role of the Pauli exclusion
principle and the Fermi-Bose mapping theorem of Girardeau applies
\cite{Girardeau1960}. This theorem proves that the wavefunction of
the system defined by a Hamiltonian such as
Eq.~\eqref{eq:hamiltonian} with $g\rightarrow\infty$ is equivalent
to the properly symmetrised wavefunction of a gas of noninteracting
fermions in the same trapping potential $V_0(x)$. As is customary
for non-interacting fermions with periodic boundary conditions, an
applied commensurate lattice $V_l(x)$ leads to the opening of a
single particle band gap of width $\triangle=V_l/4$. This is the
Mott insulating phase.

As we will focus on the pinning transition in the Tonks Girardeau
limit, let us briefly review the Fermi-Bose mapping theorem. The
essential idea is that one can treat the interaction term in
Eq.~(\ref{eq:hamiltonian}) by replacing it with a boundary condition
on the allowed manybody bosonic wave-function
\begin{equation}
  \label{eq:constraint}
  \Psi_B(x_1,x_2,\dots,x_n)=0\quad \mbox{if} \quad |x_i-x_j|=0\;,
\end{equation}
for $i\neq j$ and $1\leq i\leq\ j\leq N$. This is a hard core
constraint meaning no probability exists for two particles ever to
be at the same point in space.

This constraint is automatically fulfilled by the corresponding
noninteracting fermionic system using a Slater determinant such that
\begin{equation}
  \Psi_F(x_1,x_2,\dots,x_N) =\frac{1}{\sqrt N!}
                             \det_{n,j=1}^{N}[\psi_n(x_j)]\;,
\label{eq:psiF}
\end{equation}
where the $\psi_n(x)$ are the single particle eigenstates of the
noninteracting system in trapping potential $V_0(x)$. This, however,
leads to a fermionic rather than bosonic symmetry, which can be
corrected by a multiplication with the appropriate unit
antisymmetric function \cite{Girardeau1960}
\begin{equation}
 \Psi_B(x_1,x_2,\dots,x_N) = \prod_{1\leq i < j\leq N} \mbox{sgn}(x_i-x_j)\Psi_F(x_1,x_2,\dots,x_N)\;,
  \label{eq:mapFB}
\end{equation}
The power of the mapping theorem is that certain important many-body
quantities of the Tonks-Girardeau gas in an arbitrary external
potential, can now be calculated using single particle states. The
analytic nature of the many-body states of the gas in this limit are
convenient to explore the properties of the pinning transition.

%
%
A feature of the pinning quantum phase transition is that even a
weak lattice can change the energy spectrum dramatically and the
overlap of two ground states decreases. Using FB mapping, the ground
state fidelity can be expressed via single particle basis
\cite{Lelas}
\begin{eqnarray}
|\langle\Psi_0|\Phi_0\rangle|^2 & = & |\frac{1}{N!} \int dx_1 \cdots
dx_N \sum_{\sigma_1}(-)^{\sigma_1} \prod_{i=1}^N
\psi^{*}_{\sigma_1(i)}(x_i)
\sum_{\sigma_2}(-)^{\sigma_2} \prod_{j=1}^N \phi_{\sigma_2(j)}(x_j)|^2 \nonumber \\
& = & |\frac{1}{N!}
\sum_{\sigma_1}\sum_{\sigma_2}(-)^{\sigma_1}(-)^{\sigma_2}
\prod_{i=1}^N \int\psi_{\sigma_1(i)}^{*}(x)\phi_{\sigma_2(i)}(x)dx|^2 \nonumber \\
& = & |\det{\bf A}|^2 \label{FTG}
\end{eqnarray}
where elements of matrix $\bf{A}$ are
\begin{equation}
A_{ij}=\int\psi_i^{*}(x)\phi_j(x)dx.\label{matrixA}
\end{equation}
If the system is in the ground state $|\Psi_0\rangle$ and we
suddenly turn on optical lattice $V_l(x)$, the probability that we
will excite the system away from the initial ground state is
conveniently related to the ground state fidelity \cite{Grandi:10}
\begin{equation}
P_{exc}=1-|\langle\Psi_0|\Phi_0\rangle|^2.\label{excprob}
\end{equation}
In section \ref{sec:Loschmidt} we explore non-equilibrium dynamics
after a sudden quench of lattice amplitude.
The fidelity of the TG gas is formally equivalent to a gas of
non-interacting spin polarized fermions \cite{Goold2011}.

\section{Pinning transition for a TG gas in
an infinitely deep box: ground state fidelity} \label{sec:BOX}

In this section we apply the concept of ground state fidelity (GSF)
[Eqs. (\ref{FTG}) and (\ref{matrixA})] to study the pinning quantum
phase transition for a TG gas in an infinitely deep box
potential,
\begin{align}
V_0(x)=
\begin{cases}
0, & \mbox{if } 0\leq x\leq L \\
\infty, & \mbox{otherwise }.
\end{cases}
\label{BOX}
\end{align}
The lattice potential is defined as $V_l(x)=V_l\cos^2(kx+\phi)$. The
periodicity of the lattice corresponds to the length of the box,
$k=M\pi/L$, where $M$ is an integer. Thus, for $\phi=0$ we have
exactly $M$ wells within the box, and we expect to see the signature
of the pinning transition at $N=M$, where $N$ is the number of
particles. For the the other phases $\phi$ there are $M-1$ well
defined wells, and two half-wells at the edges of the box. In the
thermodynamic limit the differences due to boundary effects will
disappear (or become irrelevant), however in our simulations we will
investigate these finite size effects which can be relevant for
experiments. First we study the GSF numerically as a function of
number of particles $N$ for different lattice amplitudes $V_l$ and
different system sizes $L$.

\subsection{Numerical simulations}

In our numerical simulations the $x$-space grid is in units
$x_0=1\mu m$. The lattice amplitude $V_l$ and all other energies are
in units of the recoil energy $E_R=(\hbar k)^2/(2m)$. The mass $m$
corresponds to rubidium atoms $^{87}$Rb. We shall fix the lattice
wave vector to be $k=4\pi x_0^{-1}/3$ ($\lambda=2\pi/k=3x_0/2$), and
keep it constant throughout this section. The length of the box
$L=M\pi/k=M\lambda/2$ will vary. In all simulations $V_l\leq E_R$,
i.e., we are in the weak lattice regime \cite{buchler, Nagerl:10}.
Single particle (SP) states of $V_0(x)$ are
$\psi_n(x)=\sqrt{2/L}\sin(n\pi x/L)$ ($n=1,2,3,\ldots$). The SP
states $\phi_n(x)$ of $V_0(x)+V_l(x)$ are calculated numerically.
From these one obtains GSF via Eq. (\ref{FTG}).

\begin{figure}[!h]\centering
\includegraphics[scale=0.40]{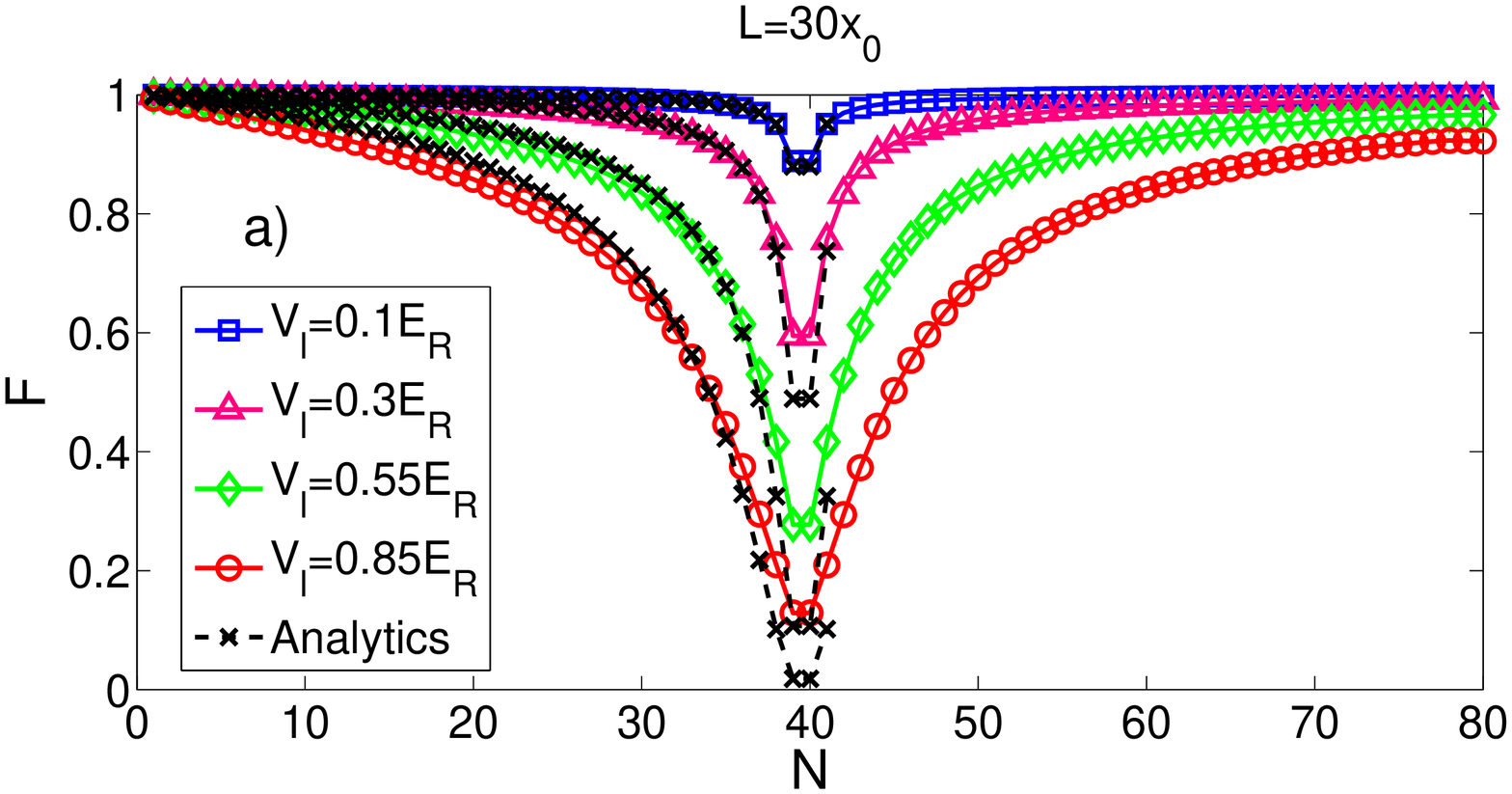}
\includegraphics[scale=0.40]{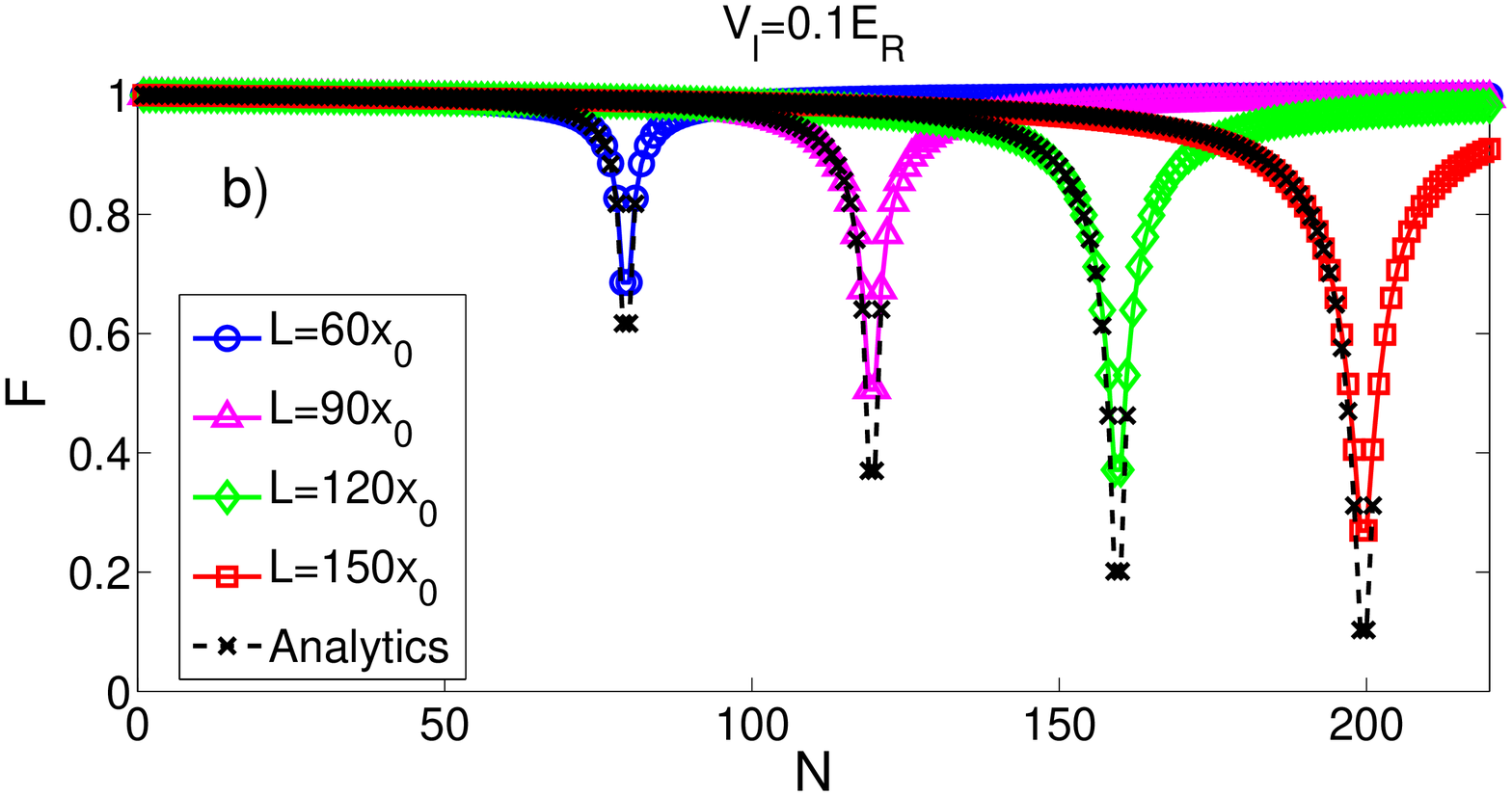}
\caption{Ground state fidelity (a) as a function of the number of
particles $N$, for constant box size $L=30x_0$ and lattice wave
vector $k=4\pi x_0^{-1}/3$ with different lattice amplitudes $V_l$.
The black crosses are analytical results (for $N\leq M+1$) obtained
with first order perturbation theory. Ground state fidelity (b) as a
function of $N$, for constant lattice amplitude $V_l=0.3E_R$ and
wave vector $k=4\pi/3$ with different box sizes $L$. See text for
details.} \label{fig1}
\end{figure}

Figure~\ref{fig1}(a) shows the GSF as a function of the number of
particles $N$, for different values of the lattice amplitude. The
phase $\phi=0$, i.e., $V_l(x)=V_l\cos^2(kx)$. The size of the box is
$L=40\pi/k=30x_0$, that is, $M=40$ and we expect the pinning to
occur at $N=40$. Indeed we observe a dramatic decrease of fidelity
when approaching commensurability, however, the GSF is equal for
$N=39$ and $N=40$. One can argue that in the thermodynamic limit
there is a single point at which the pinning takes place, and that
this anomaly is a consequence of finite size effects. Nevertheless,
such finite size effects are important for experimental systems as
they occur at the relevant densities.
The aforementioned anomaly will be explained in the next subsection
using first order perturbation theory. We point out that the GSF
obtained with the first order perturbation theory (black crosses)
for $N\leq M+1$, developed in Subsection~\ref{BOXanalysis}, is in
perfect agreement with numerics for small amplitude $V_l=0.1E_R$
(blue circles in Fig. \ref{fig1}(a)), for higher amplitudes there
are discrepancies between first order perturbation theory and exact
numerics in the dip of GSF, while outside of the GSF dip agreement
is fairly good for all amplitudes, see Fig.~\ref{fig1}(a).

Figure~\ref{fig1}(b) shows the GSF as a function of the the number
of particles $N$, for different values of $L$ (lattice amplitude is
held constant at the value $V_l=0.3E_R$). We clearly see that GSF
decreases in the region of criticality with the increase of $L$, as
expected. In the next subsection we will show that at the critical
point $F\rightarrow 0$ as $L\rightarrow \infty$.

\begin{figure}[!h]
\centering
\includegraphics[scale=0.40]{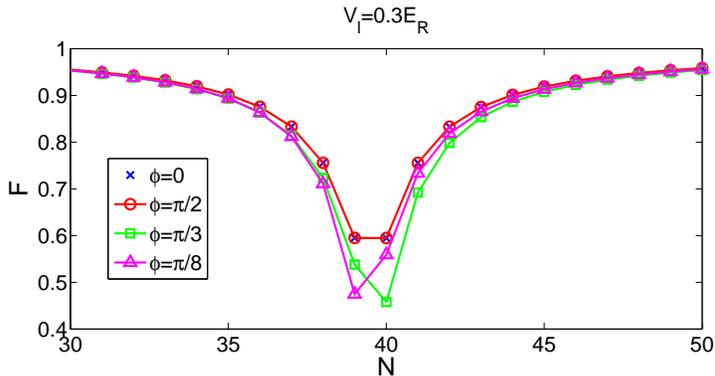}
\caption{Ground state fidelity as a function of $N$, for different
phases $\phi$ of the lattice $V_l(x)$. Outside presented interval of
$N$, GSF is approximately identical for all phases. See text for
details} \label{fig2}
\end{figure}

Let us discuss the boundary effects for a finite-size system.
Interestingly, if we use the phase $\phi=\pi/2$ for the lattice,
such that $V_l(x)=V_l\sin^2(kx)$, we obtain approximately identical
values for the fidelity. In Fig. \ref{fig2} we show GSF for cosine
squared ($\phi=0$ red circles) and sine squared ($\phi=\pi/2$ blue
crosses) lattice, these values overlap and come in pairs. This
symmetry is lost for phases $\phi$ in between $0$ and $\pi/2$. As an
example, Fig. \ref{fig2} shows GSF as a function of the number of
particles for $\phi=\pi/3$ (green squares) and $\phi=\pi/8$ (pink
triangles); there is a single point at which GSF has a minimum,
either at $N=39$ or at $N=40$. For other phases, in between $\phi=0$
and $\phi=\pi/2$, results are qualitatively and quantitatively
similar as for $\phi=\pi/3$ and $\phi=\pi/8$.

It is interesting that for a cosine squared lattice with exactly $M$
wells the signature of the pinning occurs at $N=M,M-1$, and that the
same values are obtained for a sine squared lattice with $M-1$ wells
plus two half-wells. In order to see the differences between
different boundary conditions we need to look at another quantity
rather than the GSF. We choose to investigate the behavior of the
energy (following the experiment \cite{Nagerl:10}), and the
single-particle density.

\begin{figure}[!h]
\centering
\includegraphics[scale=0.40]{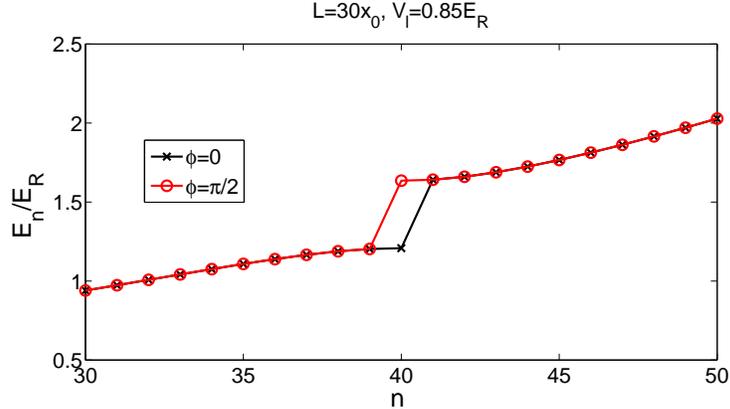}
\caption{Single particle SP energy spectrum for cosine-squared
$\phi=0$ (black crosses) and sine-squared $\phi=\pi/2$ (red circles)
lattice, other parameters are $L=30x_0$ and $V_l=0.85E_R$. See text
for details.} \label{fig3}
\end{figure}

In Fig.~\ref{fig3} we plot the single-particle (SP) energy spectrum
of the potential potential $V_0(x)+V_l(x)$ for both sine- and
cosine-squared lattice. As in Fig.~\ref{fig1}(a) the parameters are
$L=30x_0$ and $k=4\pi x_0^{-1}/3$ (which gives $M=40$), and the
lattice amplitude is $V_l=0.85E_R$. We see that the energy spectrum
is different for these two lattices. Even though we cannot strictly
speak about a gap for a finite size lattice, by observing Fig.~\ref{fig3} 
we see that the gap-like opening in the spectrum occurs
at $n=40$ ($n$ is the index of a single-particle state) for a
cosine-squared lattice, whereas for the sine-squared lattice it
occurs at $n=39$. These signatures for the pinning transition are
intuitively expected when we think of the number of particles versus
the number of wells in these two lattices. Even though we speak here
about the SP spectrum, the energy gap will be present in many-body
excitations of the TG gas as well, because of the FB mapping
\cite{Girardeau1960}. We emphasize that 'gap' in SP spectrum occurs
for SP states with the same (or approximately the same) wavelength
as the lattice wavelength; this fact will be used in Section
\ref{sec:HO}.



\begin{figure}[!h]
\centering
\includegraphics[scale=0.40]{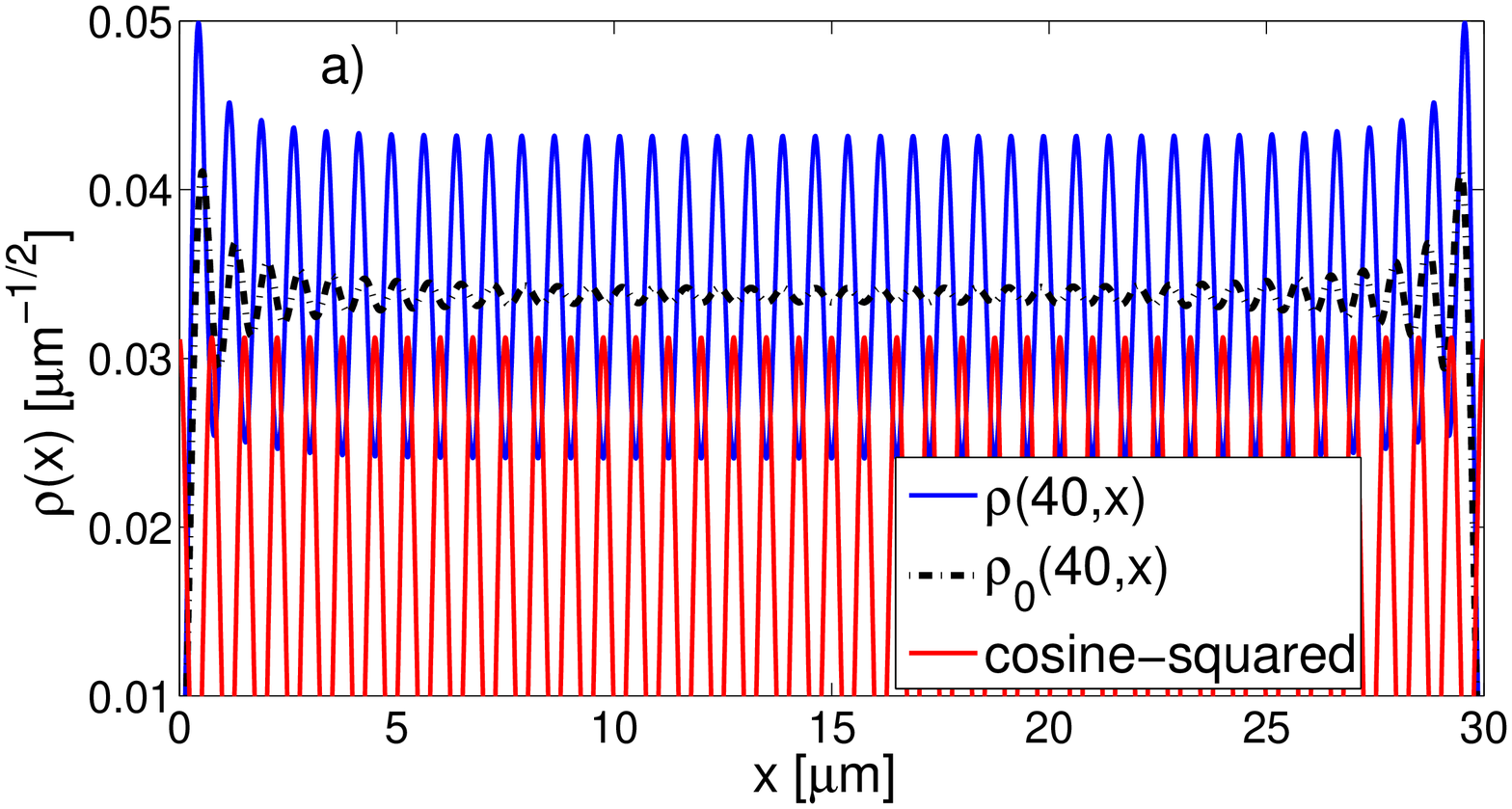}
\includegraphics[scale=0.40]{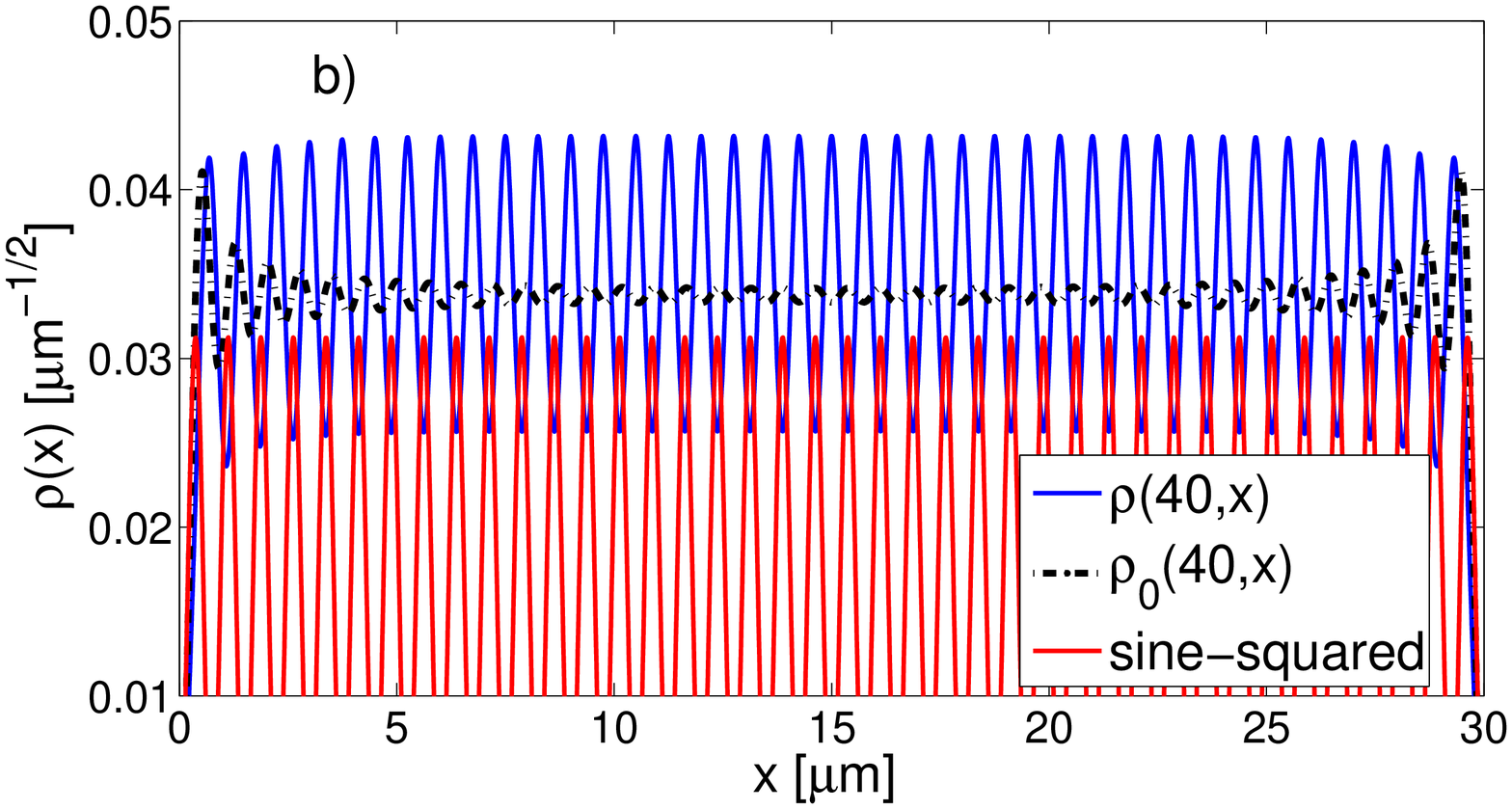}
\caption{Single particle density $\rho(N,x)$ obtained from the many-body
ground states of $N$ atoms in $V_0(x)+V_l(x)$ potential (a)
$\rho(40,x)$ (blue line) in cosine-squared lattice (red line) and
(b) $\rho(40,x)$ (blue line) in sine-squared lattice (red line). For
reference we plot the corresponding densities $\rho_0(N,x)$ of $N$ atoms
in the ground state of $V_0(x)$ trap (black doted line). See text for
details.} \label{fig4}
\end{figure}

We now turn to the single-particle densities
$\rho(N,x)=\sum_{n=1}^N|\phi_n(x)|^2/N$, which are plotted in
Fig.~\ref{fig4} for the cosine- (a) and the sine-squared (b)
lattice, in comparison with the density
$\rho_0(N,x)=\sum_{n=1}^N|\psi_n(x)|^2/N$, and the lattice maxima
and minima (parameters are $N=M=40$, $V_l=0.55E_R$ and $L=30x_0$).
We see that in both cases the density maxima occur at the lattice
minima as expected; the two cases differ at the boundary which is
reflected in the energy spectrum but not in the GSF.

\begin{figure}[!h]
\centering
\includegraphics[scale=0.40]{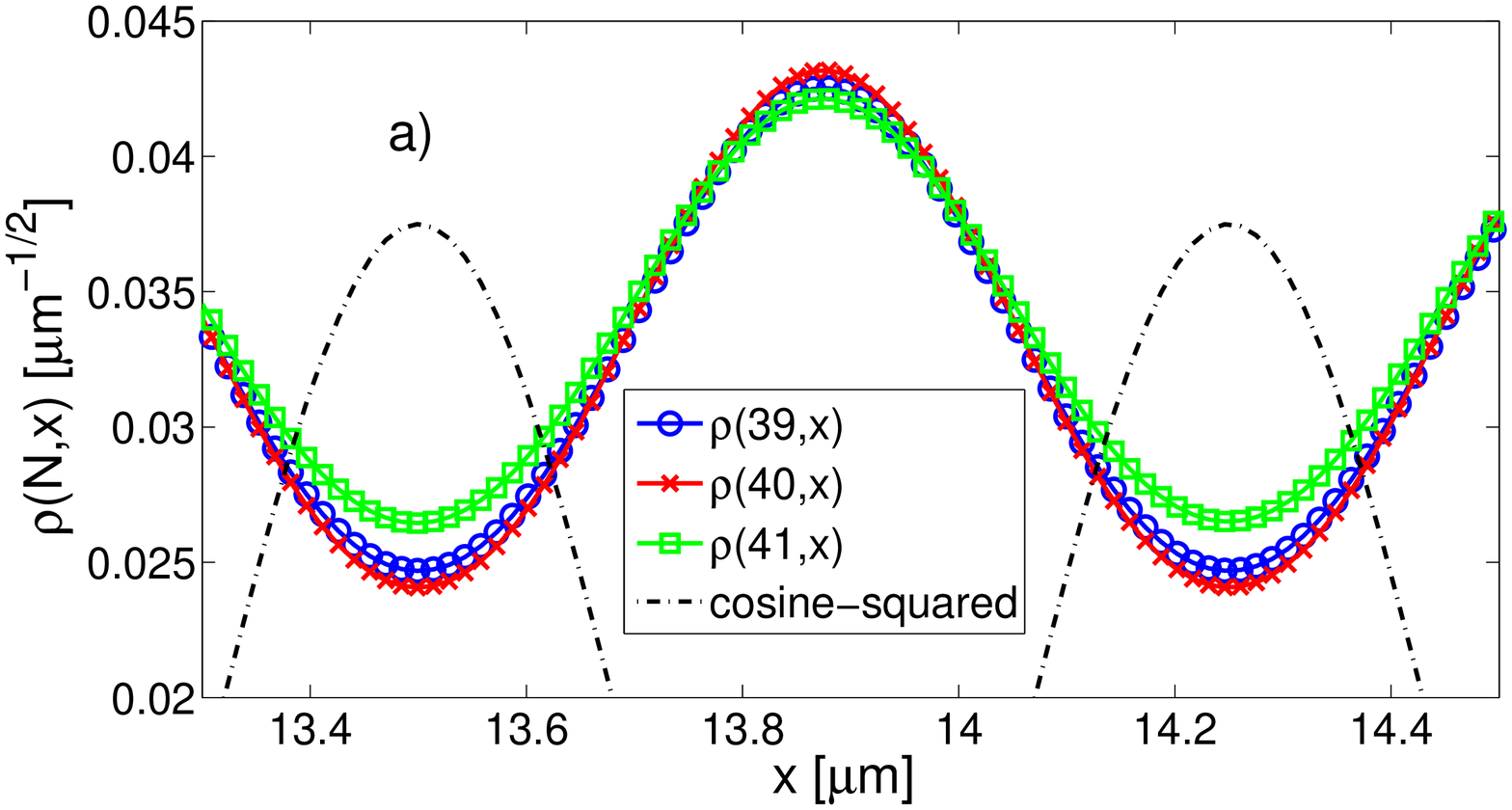}
\includegraphics[scale=0.40]{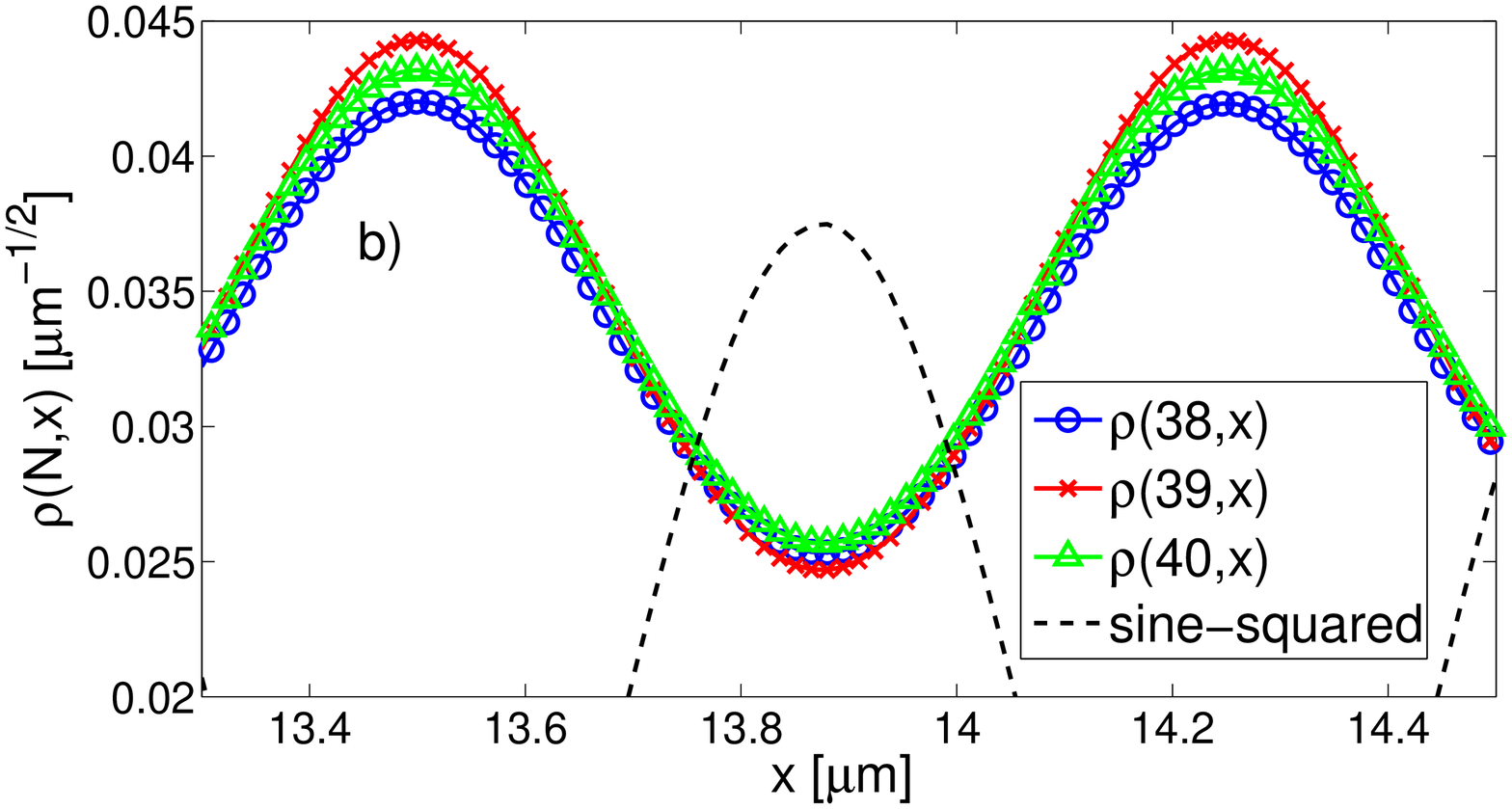}
\caption{Single particle density. (a)  $\rho(40,x)$ (red cross) and
two nearby densities $\rho(39,x)$ (blue circles) and $\rho(41,x)$
(green squares) in cosine lattice (blue doted line). (b)
$\rho(39,x)$ (red crosses) and two nearby densities $\rho(38,x)$
(blue circles) and $\rho(40,x)$ (green triangles) in sine lattice
(blue doted line). A smaller range of axis is chosen to provide good
visibility. See text for details.} \label{fig5}
\end{figure}

In~Fig.~\ref{fig5} (a) we show the inset of the density $\rho(N=M,x)$
(red crosses) vs. two nearby densities $\rho(N=M-1,x)$ (blue
circles) and $\rho(N=M+1,x)$ (green squares) in the cosine-squared
lattice (blue doted line). We clearly see that the probability for
particles to 'sit' at the minima of the cosine-squared lattice is
the highest for $N=M$ atoms (also, the probability for the atoms to
sit at potential maxima is the lowest for $N=M$ atoms). This
observation confirms the indication given by SP energy spectrum in
Fig.\ref{fig2} regarding where the pinning takes place in the finite
size system. In Fig.~\ref{fig5} (b) we show the same quantities for
the sine-squared lattice. We see that the signature of pinning is
strongest at $N=M-1$ consistent with the single-particle
spectrum.

We see that the energy and the single particle density can
distinguish between different types of boundary conditions, whereas
the GSF is less sensitive to these effects. The GSF has an advantage
over the energy spectrum in the thermodynamic limit where it
dramatically shows where the pinning takes place for an
infinitesimally small lattice amplitude.

\subsection{Analysis of GSF via 1st order perturbation theory}\label{BOXanalysis}

In this subsection we study the GSF in the context of the pinning
phase transition for the box potential via stationary first-order
perturbation theory. Unperturbed states are the SP basis of $V_0(x)$
i.e $\psi_n(x)=\sqrt{(2/L)}\sin(n\pi x/L)$. For the moment let us
focus on the cosine-squared lattice $V_l(x)=V_l\cos^2(M\pi x/L)$,
which we treat as the small stationary perturbation and denote SP
basis of $V_0(x)+V_l(x)$ as $\phi_n(x)$. To first order in the
lattice amplitude, the single-particle states of the potential
$V_0(x)+V_l(x)$ are
\begin{equation}
\phi_i(x)\propto\psi_i(x)+a_{2M-i}\psi_{2M-i}(x)+a_{2M+i}\psi_{2M+i}(x),
\label{phi1st}
\end{equation}
where $i\in \{1,\ldots,M-1,M+1,\ldots,2M-1\}$; the case $i=M$ is
treated separately. This interval of indices $i$ cover all particle
numbers of interest i.e $N=1,\ldots,2M-1$, and the criticality
region $N\sim M$ is in the center of that interval. The coefficients
are given by $a_{2M-i}=V_l/[4(E_{2M-i}-E_{i})]$ and
$a_{2M+i}=V_l/[4(E_{2M+i}-E_{i})]$ where $E_i$ is the SP energy of
$i$-th state in the $V_0(x)$ potential. Since $E_i=(i/M)^2E_R$ we
can write
\begin{equation}
a_{2M\mp i}=\frac{M}{16(M\mp i)}\frac{V_l}{E_R}.\label{coeff}
\end{equation}
The coefficient $a_{2M+i}$ in Eq. (\ref{phi1st}) can be ignored
because of the denominator in Eq. (\ref{coeff}), i.e., for $i\neq M$
\begin{equation}
\phi_i(x)\approx
\frac{\psi_i(x)+a_{2M-i}\psi_{2M-i}(x)}{\sqrt{1+|a_{2M-i}|^2}},\label{phi}
\end{equation}
where we have normalized the wave function. We see that perturbation
will be most dominant when $i=M-1$ and $i=M+1$.

For $i=M$, the first-order perturbation theory gives
$\phi_M(x)\propto\psi_M(x)+a_{3M}\psi_{3M}(x)$, where
$a_{3M}=V_l/[4(E_{3M}-E_{M})]=V_l/32E_R$ is sufficiently small for a
weak lattice $V_l/E_R\ll 1$, and we can write
\begin{equation}
\phi_M(x)\approx \psi_M(x).\label{phiM}
\end{equation}
In fact, this relation will hold even for deeper lattices as long as
$V_l/32E_R\ll 1$.

In order to calculate the ground state fidelity $F=|\det{\bf A}|^2$
we need to evaluate the matrix elements
$A_{ij}=\int\psi_i^*(x)\phi_j(x)dx$ [see Eqs. (\ref{FTG}) and
(\ref{matrixA})]. We first consider the case $N<M$. We use Eqs.
(\ref{phi}) and (\ref{phiM}) to get matrix elements $A_{ij}$ within
first order perturbation theory:
\begin{equation}
A_{ij}\approx\frac{\delta_{ij}+a_{2M-j}\delta_{i,2M-j}}{\sqrt{1+|a_{2M-j}|^2}},\label{matrixA2}
\end{equation}
where $i,j=1,\ldots,N$. If $N<M$, then the second delta term in
(\ref{matrixA2}) is zero, and the matrix (\ref{matrixA2}) is
diagonal $A_{ii}=(\sqrt{1+|a_{2M-i}|^2})^{-1}$. Thus, the GSF
($N<M$) is
\begin{equation}
F \approx\prod_{i=1}^{N}
|A_{ii}|^2\approx\prod_{i=1}^N\frac{1}{1+|a_{2M-i}|^2}\label{FTG1}
\end{equation}
Since the coefficients $|a_{2M-i}|^2$ rise quadratically as $i$
approaches $M$, we understand the behavior of GSF when $N$
approaches $M$ from below, which was observed numerically in Fig.
\ref{fig1} (a) and (b). In Fig. \ref{fig1}(a) we plot results of
Equation (\ref{FTG1}) (black crosses), agreement with exact
numerical results is excellent for small amplitudes, i.e,
$V_l=0.1E_R$ in Fig. \ref{fig1}(a) (blue squares), for larger
amplitudes agreement is good outside the dip of GSF, while we see
discrepancies in the dip, resulting from first order perturbation
theory are systematically lower than exact numerics.

The case $N=M$ is straightforward due to $A_{MM}\approx 1$, and for
$N=M$ the GSF becomes
\begin{equation}
F \approx
\prod_{i=1}^{M-1}\frac{1}{1+|a_{2M-i}|^2}|A_{MM}|^2,\label{Fdip}
\end{equation}
which is identical to the value for $N=M-1$ (see Eq. (\ref{FTG1})),
which explains our numerical observation. This is an interesting
observation. The GSF will decrease when first order perturbation is
the most effective. We expect it to be the most effective at
commensurability, $N=M$. However, the coefficient at $N=M-1$
contributes the most in this sense, whereas for $N=M$ the
perturbation on the SP eigenstates (which is reflected onto the many
body eigenstates via FB mapping) is essentially negligible.

If we enlarge the box to new size e.g $L'=2L$, and we keep the
lattice wave vector constant $k=M\pi/L$, we have $M'=2M$, and the
fidelity dip moves to $N=2M-1$ and $N=2M$, and decreases in value
because the new coefficient $a_{M'}=2a_M$, which gives smaller
product terms in equation (\ref{FTG1}). This explains results of Fig
\ref{fig1}(b) where we vary system size $L$.

The first-order perturbation theory also provides an explanation for
the influence of the phase $\phi$. For example, for the sine-squared
lattice, because $\cos^2(kx)=1-\sin^2(kx)$, the integrals appearing
in the perturbation expansion are
$$\int\psi_i^{*}(x)\cos^2(kx)\psi_j(x)dx=-\int\psi_i^{*}(x)\sin^2(kx)\psi_j(x)dx,$$
and the coefficients change sign, wave functions differ, but the GSF
(\ref{FTG1}) depends on absolute squares of these coefficients and
is insensitive to this phase. This explains results of Fig.
\ref{fig2} for cosine-squared and sine-squared lattice (red circles
and blue crosses). For some arbitrary phase value between $0$ and
$\pi/2$, the main difference is that equations (\ref{phi1st}) and
(\ref{phiM}) will no longer hold and more coefficients are needed in
expansion of $\phi_i(x)$ in terms of $\psi_i(x)$, especially for
$i\sim M$, which breaks the symmetry between $N=M$ and $N=M-1$
cases.


Let us finally discuss the $N=M+1$ case for the cosine-squared
lattice. Matrix ${\bf A}$ acquires two off diagonal elements
$A_{M+1,M-1}$ and $A_{M-1,M+1}$ with the following property
\begin{equation}
A_{M+1,M-1}=-A_{M-1,M+1}\approx\frac{a_{M+1}}{\sqrt{1+|a_{M+1}|^2}},\label{property}
\end{equation}
due to Eq. (\ref{coeff}). In this case the determinant of matrix
(\ref{matrixA2}) becomes
$$\det{\bf A}\approx\prod_{i=1}^{M-2}A_{ii}A_{MM}(A_{M-1,M-1}A_{M+1,M+1}-A_{M+1,M-1}A_{M-1,M+1}).$$
Due to (\ref{coeff}) and (\ref{property}) we have
$(A_{M-1,M-1}A_{M+1,M+1}-A_{M+1,M-1}A_{M-1,M+1})\approx1$, and since
$A_{MM}\approx 1$, we finally get that determinant of matrix
(\ref{matrixA2}) for $M+1$ particles is
$$\det{\bf A}\approx\prod_{i=1}^{M-2}A_{ii}.$$
From the last relation we see that the ground state fidelities for
$N=M+1$ and $N=M-2$ particles are approximately equal (see Eq.
(\ref{FTG1})). In Fig. \ref{fig1}(a) we plot these results for
$N=(M,M+1)$ particles (black crosses) in addition to GSF for $N<M$.
One could proceed to other values of $N>M+1$ in the same
fashion and analyze the GSF via perturbation theory. 

\section{The Pinning transition of the Tonks-Girardeau gas in the harmonic oscillator:
ground state fidelity} \label{sec:HO}

In this section we study the pinning transition in the
experimentally relevant harmonic oscillator (HO) potential
\cite{Nagerl:10}
\begin{equation}
V_0(x)=\frac{m\omega_0^2x^2}{2}.\label{HO}
\end{equation}
We choose parameters following the experiment in Ref.
\cite{Nagerl:10}, i.e., the atoms inside the trap are caesium atoms,
$^{133}$Cs. The optical lattice is $V_l(x)=V_l\sin^2(kx)$ with
$k=1.88\pi x_0^{-1}$ where $x_0=1\mu$m and the wavelength of the
lattice is $\lambda=1064$nm. Lattice amplitude and all other
energies are in units of the recoil energy $E_R=(\hbar k)^2/(2m)$.

\begin{figure}[!h] \centering
\includegraphics[scale=0.38]{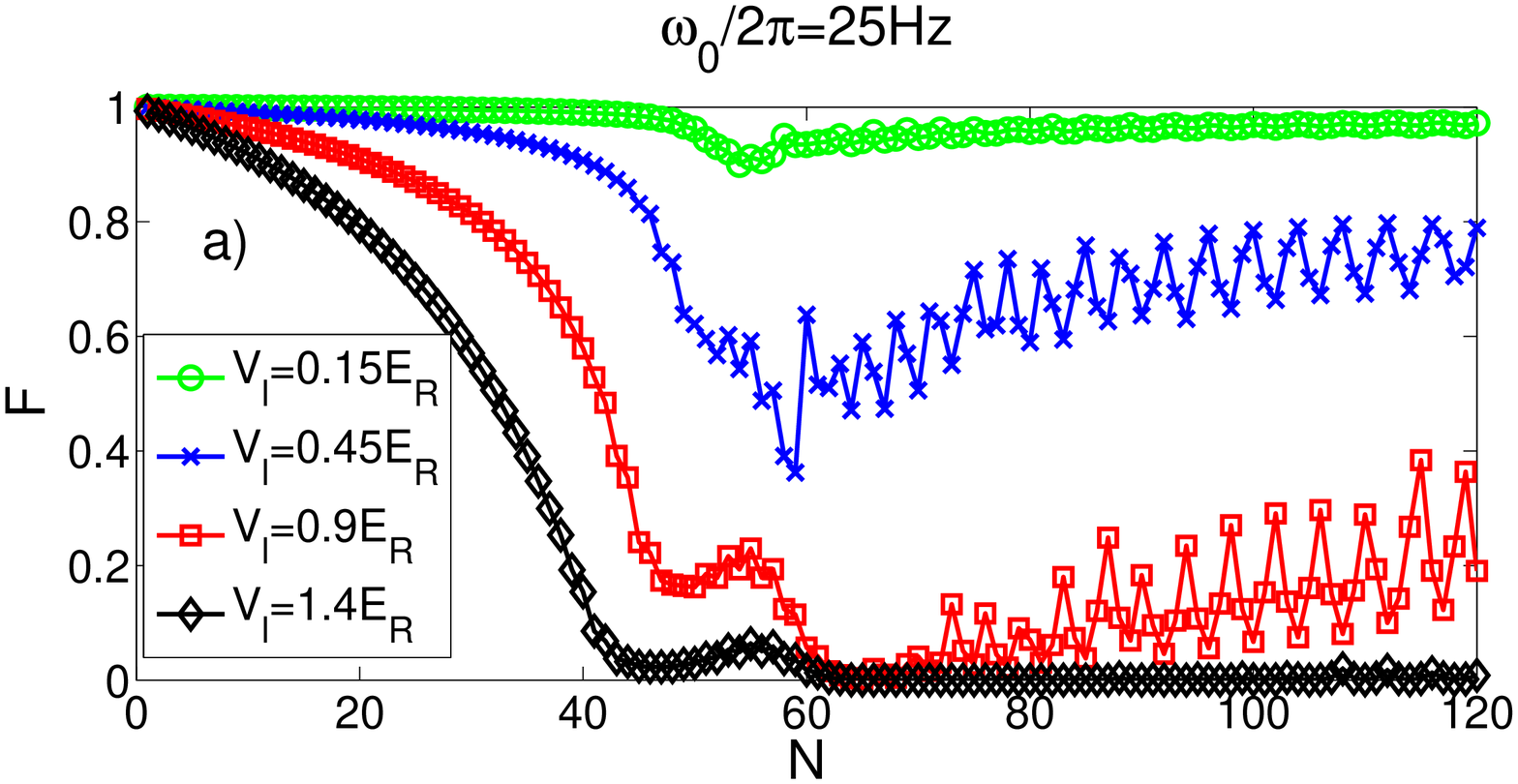}
\includegraphics[scale=0.38]{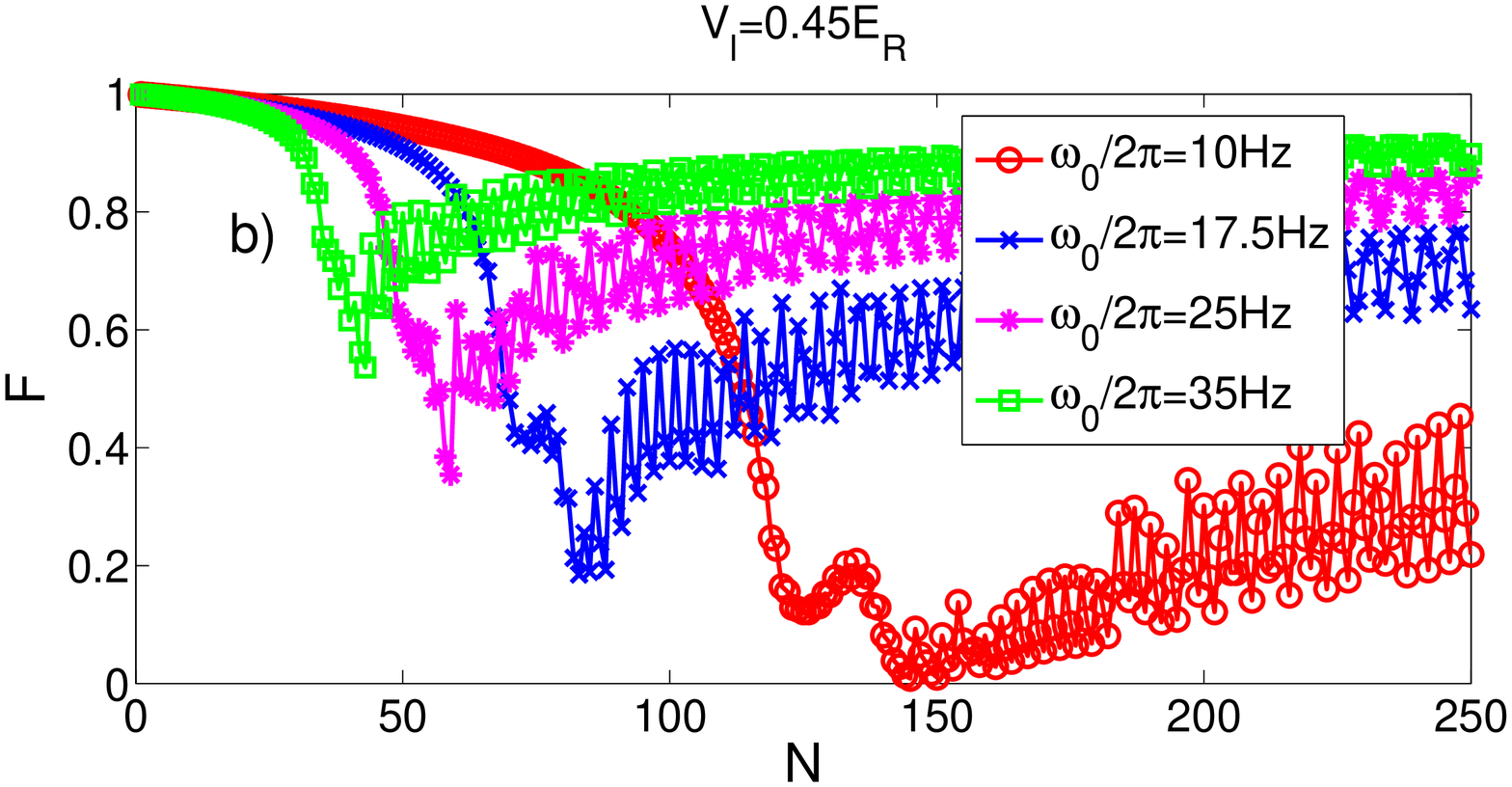}
\caption{Ground state fidelity for the pinning transition in a
harmonic oscillator potential. (a) GSF as a function of number of
particles $N$ for different lattice amplitudes $V_l$ with constant
frequency $\omega_0/2\pi=25$Hz. (b) GSF as a function of $N$ for
different $\omega_0$ with constant $V_l=0.45E_R$. See text for
details.} \label{fig6}
\end{figure}

In Fig.~\ref{fig6}(a) we plot the ground state fidelity as a
function of number of particles for different values of the lattice
amplitude $V_l$. The harmonic oscillator frequency is
$\omega_0/2\pi=25$Hz (similar to the frequency used in experiment
$\omega_{exp}/2\pi=22(3)$Hz \cite{Nagerl:10}). We see for all
amplitudes $V_l$ that GSF first decreases smoothly, in a similar
fashion as in the infinitely deep box in Fig. \ref{fig1}(a) and (b).
Then GSF decreases faster until it reaches first minimum at
$N\approx50$ after which it develops oscillations with deeper
minima. For smaller amplitudes $V_l=0.15E_R$ (green circles) and
$V_l=0.45E_R$ (blue x's) the oscillations have a global minimum at
$N\approx60$, and after that the GSF starts to rise, as expected,
but at a slower rate compared to the rate of decrease towards the
first minimum, in contrast to the square well. For higher amplitude
$V_l=0.9E_R$(red squares) GSF is effectively zero i.e $F\approx0$
for interval of $N$'s between $N\sim60$ and $N\sim70$, with the slow
increase of average GSF for $N$ above $70$. Finally, for still
higher amplitude $V_l=1.4E_R$ (black diamonds) (similar to the amplitude
used in experiment $V_{exp}=1.5(1)E_R$ \cite{Nagerl:10}) GSF drops
to values slightly above zero already for $N\approx45$ and after a
small bump we see that $F\approx0$ from $N\sim60$ to $N\sim110$;
above $N\sim110$ GSF slowly rises and develops oscillations (not
shown) similar to GSF for $V_l=0.9E_R$(red squares).

%
%

In Fig.~\ref{fig6}(b) we plot the ground state fidelity $F$ as a
function of $N$ for different values of the harmonic oscillator
frequency $\omega_0$ with constant lattice amplitude $V_l=0.45E_R$.
We see, as expected, that the pinning transition occurs for larger
$N$, and the fidelity dip lowers.

We point out that the results of Fig.~\ref{fig6}(a), obtained for
$\gamma\gg1$, are in excellent agreement with the experimental results
of Ref. \cite{Nagerl:10}, obtained for large but finite $\gamma$,
where it is said that commensurability of superfluid phase and the
lattice is best fulfilled when there are about $N\sim60$ atoms in
the central tube. We see in Fig. \ref{fig6}(a) for all amplitudes
GSF shows enhanced sensitivity and strongest decay in  the region
$N\sim60$.

In order to understand these results, we need to define the
commensurability of the Tonks-Girardeau gas and the applied optical
lattice. This is not straightforward because of the inhomogeneous
atomic density in the ground state of the harmonic oscillator
potential. We draw upon the results of Section \ref{sec:BOX}, where
the GSF had a minimal value when the unperturbed $N$th SP state
entering the $N$-particle ground state via Eq. (\ref{eq:mapFB}), had
the same wavelength as the optical lattice. In the case of the
harmonic trap (\ref{HO}), the asymptotic expansion of SP states
$\psi_n(x)$ for $n\gg1$ is
\begin{equation}
\psi_n(x/a_0)\propto\cos(\sqrt{2n}x/a_0-n\pi/2),\label{asymp}
\end{equation}
where $a_0=\sqrt{\hbar/m\omega_0}$. This provides us with the
dominant wavelength of the $n$-th SP state. We estimate that the
pinning occurs when
\begin{equation}
k\approx \sqrt{2N}/a_0,
\end{equation}
which yields
\begin{equation}
N_{pinn}\approx\frac{k^2\hbar}{2m\omega_0}\label{NHO}
\end{equation}
for the number of particles where pinning occurs. Eq.~(\ref{NHO}) is
obtained for $n\gg1$; in experiments one usually has $N>30$. In
addition, we stress that Eq.~(\ref{NHO}) is in agreement with
Ref.~\cite{buchler}, where the pinning transition is said to occur
for such $N$ that the peak density of the superfluid phase, obtained
with Thomas Fermi approximation for $\gamma\gg 1$, is equal to the
commensurate density $n_c=2/\lambda$. For $\omega/2\pi=25$Hz our
estimate yeilds $N_{pinn}\sim50$, which explains the drop in the GSF
observed in Fig. \ref{fig6}(a). Equation (\ref{NHO}) also explains
the positions of the first minima of GSF in Fig. \ref{fig6}(b) since
it gives $N_{pinn}\sim\{38,52,75,130\}$ for
$\omega_0/2\pi=\{35,25,17.5,10\}$Hz, respectively, in fair agreement
with exact numerical results. Again, as we make the system larger,
the minimum value of GSF decreases which is consistent with the
decrease of GSF at criticality in the thermodynamic limit.

\begin{figure}[!h] \centering
\includegraphics[scale=0.4]{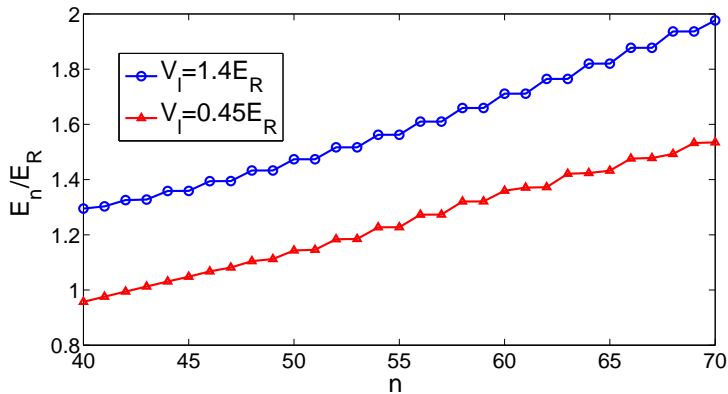}
\caption{Single particle energy spectrum of harmonic trap with the
lattice $V_0(x)+V_l(x)$, for two values of amplitude $V_l=1.4E_R$
(blue circles) and $V_l=0.45E_R$ (red triangles). See text for
details.} \label{fig7}
\end{figure}

Besides the GSF, it is instructive also to look into the single
particle energy spectrum  plotted in Fig. \ref{fig7} for
$V_0(x)+V_l(x)$ with the amplitudes $V_l=0.45E_R$ (red triangles)
and $V_l=1.4E_R$ (blue circles). We see that for $V_l=0.45E_R$ (red
triangles) at $n\sim50$ and larger values a series of 'gaps' open up
in the sense that at some $N$ values the excitations of the TG gas
from the ground state cost more energy.

These results can also be understood simply in terms of
commensurability of the SP density of the TG gas in the HO potential and the
lattice $V_l(x)$. SP density of $N$ particles in the ground state of
TG gas in $V_0(x)$ potential is
$\rho_0(N,x)=\sum_{n=0}^{N-1}|\psi_n(x)|^2$. This function is
inhomogeneous with maximum value (peak density) at $x=0$ (we now
ignore the small auxillary oscillations which can lead to the local
minimum for $x=0$ depending on the parity of $n=N-1$th state). When we
increase $N$ (starting from $N=1$), the first part of the density
$\rho_0(N,x)$ that approaches the commensurability condition
$n_c=2/\lambda$, is the central part i.e for $\rho_0(N,0)$. Thus the
first 'gap', and first minima of GSF, will appear for $N_{pinn}$
with property $\rho_0(N_{pinn},0)\approx n_c$, in accordance with
Ref. \cite{buchler} and Equation (\ref{NHO}). Adding more particles,
i.e $N>N_{pinn}$, leads to $\rho_0(N,0)>n_c$ but now in some regions
left and right from $x=0$, the density becomes commensurate with
lattice i.e $\rho_0(N,-d)\approx n_c$ and $\rho_0(N,d)\approx n_c$,
for some $d>0$, and pinning still occurs, i.e, additional 'gaps' in
SP spectrum are present and GSF still lowers. The distance $d$
increases with the increase of $N$ and commensurability condition is
satisfied for regions further towards the edges of the trap
$V_0(x)$. The fraction of the atomic cloud commensurate with the
lattice gets smaller and as a consequence the GSF slowly increases.
Oscillations are present due to many fine details such as the
interplay of symmetry of the lattice and the symmetry of the
$\psi_N(x)$ states.

%


\section{The Loschmidt echo and out of equilibrium dynamics: sudden quench with optical lattice}
\label{sec:Loschmidt}

In this section we look for signatures of the pinning transition in
the non-equilibrium dynamics of TG gas, initially in ground state of
some trapping potential $V_0(x)$, after optical lattice $V_l(x)$ is
suddenly turned on. This is an example of a sudden 'quench'. Before
the quench with the lattice potential, the gas is in the equilibrium
ground state $|\Psi_0\rangle$ of Hamiltonian $\hat H_{0}$, see
(\ref{eq:hamiltonian}). At $t=0$ we suddenly turn on the lattice
potential $V_l(x)$ and an out-of-equilibrium many body state
$|\Phi(t)\rangle=\exp[-i(\hat{H_0}+V_l(x))t/\hbar]|\Psi_0\rangle$
starts to evolve, where $|\Phi(0)\rangle=|\Psi_0\rangle$ is the
initial condition. We would like to develop a quantitative
understanding of the out of equilibrium dynamics which is encoded in
this state. Conveniently, the mapping theorem also holds for time
dependent states, and the quenched state $|\Phi(t)\rangle$ can be
constructed using a Slater determinant of time evolving single
particle states such that $|\Phi(t)\rangle = \frac{1}{\sqrt
N!}\prod_{1\leq i < j\leq N}
\mbox{sgn}(x_i-x_j)\det_{n,j=1}^{N}\left[\psi_n(x_j,t)\right]$. The
single particle states $\psi_n(x_j,t)$ are out of equilibrium, and
obtained by solving
$i\hbar\partial_t\psi_n(x,t)=\left[-\hbar^{2}/(2m)\partial_x^2+
V_0(x)+V_l(x)\right]\psi_n(x,t)$ with initial conditions
$\psi_n(x,0)=\psi_n(x)$ where $\psi_n(x)$ are the initial single
particle states (SP) which are used to construct the unperturbed
ground state $|\Psi_0\rangle$ i.e SP states of $V_0(x)$ potential.

A prototypical quantity to calculate for a system perturbed out of
equilibrium is the so-called Loschmidt echo \cite{Gorin}, which is
defined as
$$L(t)=|\langle\Psi_0|\exp(i\hat{H}_0t/\hbar)\exp[-i(\hat{H}_0+V_l(x))t/\hbar]|\Psi_0\rangle|^2.$$
It is a measure of the sensitivity of the system to the quench
protocol, which in this case is simply the application of the
external lattice potential to the initial Tonks gas equilibrium
state. Despite it's mathematical simplicity, it conveys a great deal
of information about the many-body system under scrutiny, such as
universal behavior at criticality \cite{Zanardi:06} and important
information on the thermalization of observables. Closed formulas
for the echo are, in general, very difficult to obtain. For a
Tonks-Girardeau gas, the Loschmidt echo was recently computed in a
relatively straightforward way \cite{Lelas}.

Since $\hat{H}_0|\Psi_0\rangle=\Omega_0|\Psi_0\rangle$, where
$\Omega_0$ is ground state energy of TG gas in $V_0(x)$ trap, we get

\begin{equation}
L(t)=|\langle\Psi_0|\exp[-i(\hat{H}_0+V_l(x))t/\hbar]|\Psi_0\rangle|^2=|\langle\Psi_0|\Phi(t)\rangle|^2.\label{LE}
\end{equation}
Relation (\ref{LE}) shows that in our case the Loschmidt echo (LE)
is equivalent to the survival probability i.e probability that
system will be in the initial state at the time $t$ after the
quench. We will interchangeably use the terms Loschmidt echo and
survival probability.

The Fermi-Bose mapping theorem is valid for time dependent wave
functions thus LE can be written in a form convenient for
calculation, analogous to the calculation of the static fidelity,
\begin{eqnarray}
L(t)& = &|\frac{1}{N!} \int dx_1 \cdots dx_N
\sum_{\sigma_1}(-)^{\sigma_1} \prod_{i=1}^N
\psi^{*}_{\sigma_1(i)}(x_i,0)
\sum_{\sigma_2}(-)^{\sigma_2} \prod_{j=1}^N \psi_{\sigma_2(j)}(x_j,t)|^2 \nonumber \\
& = &|\det\mathbf{A}(t)|^2 \label{LETG}
\end{eqnarray}
and $\mathbf{A}(t)$ is the time dependent matrix containing overlaps
between static SP states of the $V_0(x)$ potential i.e $\psi_i(x,0)$ and
SP states $\psi_i(x,t)$ evolved in perturbed potential
$V_0(x)+V_l(x)$
\begin{equation}
A_{ij}(t)=\int\psi_i^{*}(x,0)\psi_j(x,t)dx. \label{matrixP}
\end{equation}
Equations (\ref{LETG}) and (\ref{matrixP}) were recently used to
study the  long time behavior of many-particle quantum decay
\cite{delCampo}. The LE of the TG gas is formally equivalent to the
corresponding echo for a gas of non-interacting fermions
\cite{Goold2011}. The Loschmidt echo of one-dimensional interacting Bose
gases was recently related \cite{Lelas} to series of experiments
\cite{Hofferberth2007,Hofferberth2008,Kruger2010} and theoretical
studies \cite{Polkolnikov2006,Gritsev2006,Bistritzer2007,
Burkov2007,Mazets2008,Stimming2011} on interference between split
parallel 1D Bose systems.

\subsection{Infinitely deep well}

In this subsection we use (\ref{LETG}) and (\ref{matrixP}) to
explore the LE of a TG gas after a sudden quench with optical lattice
$V_l(x)$, i.e at $t=0$ we suddenly turn on the lattice and leave it
on. Before the quench TG gas is in ground state $|\Psi_0\rangle$ of
infinitely deep well (\ref{BOX}) potential $V_0(x)$. In this
subsection we use same units and parameters as in section
\ref{sec:BOX}.

\begin{figure}[!h] \centering
\includegraphics[scale=0.40]{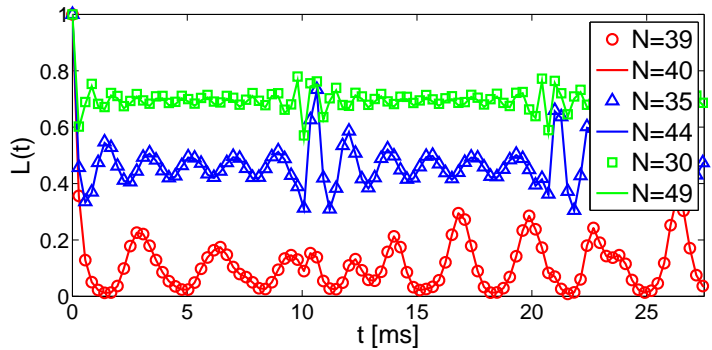}
\caption{ Loschmidt echo $L(t)$ for different particle numbers $N$
in the infinitely deep box potential. Optical lattice $V_l(x)$ has
$40$ minima, i.e, $M=40$ lattice wells. LE reflects properties of
GSF, i.e, decay is strongest for $N=(M-1, M)$, and Loschmidt echo
appears in pairs. The frequency of revivals decreases as we approach
criticality. See text for details.}\label{fig8}
\end{figure}

In Fig.~\ref{fig8} we show Loschmidt echo $L(t)$, as a function of
time for different numbers of particles $N$ obtained with exact
numerical evolution, for optical lattice $V_l(x)=0.55E_R\cos^2(4\pi
xx_0^{-1}/3)$. Size of the well $L=30x_0$. Number of lattice wells
is $M=40$. That set of parameters is the same as the one we used for
ground state fidelity $F$ in Fig. \ref{fig1}(a) denoted with red
circles. We see that properties of $F$ are reflected in LE. Values
of LE goes in pairs, i.e, curves are approximately equal for $N=M-j$
and $N=M+j-1$ particles, with $j=1,\ldots,M-1$. Decay of Loschmidt
echo is strongest and fastest close to criticality, i.e, for $N=39$
and $N=40$ particles (red circles and red solid line). In addition,
the oscillations get slower as we approach criticality.

In order to understand the results of Fig. \ref{fig8} we use relations
(\ref{coeff})-(\ref{phiM}) to write expansion of the unperturbed SP
states of $V_0(x)$ in terms of the perturbed SP states of
$V_0(x)+V_l(x)$ i.e
\begin{equation}
\psi_i(x,0)\approx\frac{\phi_i(x)-a_{2M-i}\phi_{2M-i}(x)}{\sqrt{1+|a_{2M-i}|^2}}.\label{psi}
\end{equation}
For $i=1,\ldots,M-1,M+1,\ldots,2M-1$. In case of $i=M$ we get
\begin{equation}
\psi_M(x,0)\approx\phi_M(x).\label{psiM}
\end{equation}
Using relation (\ref{psi}) we get time evolution of $\psi_j(x,t)$
after the quench
\begin{equation}
\psi_j(x,t)\approx\frac{\exp(-iE_j^lt/\hbar)\phi_j(x)-a_{2M-j}\exp(-iE_{2M-j}^lt/\hbar)\phi_{2M-j}(x)}{\sqrt{1+|a_{2M-j}|^2}},\label{psiT}
\end{equation}
where $E_n^l$ are SP energy's in $V_0(x)+V_l(x)$ potential. Now we
use (\ref{psi})-(\ref{psiT}) in Equations (\ref{matrixP}) and
(\ref{LETG}) in order to obtain insight in to the behavior of the
Loschmidt echo. The following analysis is very similar to analysis
done in subsection \ref{BOXanalysis} for the ground state fidelity.

First, we consider the case when the number of particles is less
than the number of wells, i.e $N<M$. In this case matrix
(\ref{matrixP}) is diagonal with
$$A_{ii}(t)\approx\frac{\exp(-iE_i^lt/\hbar)+|a_{2M-i}|^2\exp(-iE_{2M-i}^lt/\hbar)}{1+|a_{2M-i}|^2},$$
and since $L(t)=|\det{\bf A}(t)|^2=\det\left[{\bf A}(t){\bf
A^*}(t)\right]$ we get
\begin{equation}
\L(t)\approx\prod_{i=1}^N|A_{ii}(t)|^2,\label{LNLM}
\end{equation}
where 
\begin{equation}
|A_{ii}(t)|^2\approx\frac{1+2|a_{2M-i}|^2\cos\left[(E_{2M-i}^l-E_i^l)t/\hbar\right]+|a_{2M-i}|^4}{\left(1+|a_{2M-i}|^2\right)^2}.\label{matrixP2}
\end{equation}
If we use expression (\ref{FTG1}) for the ground state fidelity
($F$) we get for the Loscmidt echo of $N<M$ particles
\begin{equation}
\L(t)\approx
F^2\prod_{i=1}^N(1+2|a_{2M-i}|^2\cos\left[(E_{2M-i}^l-E_i^l)t/\hbar\right]+|a_{2M-i}|^4),\label{LNLM1}
\end{equation}
we see that ground state fidelity is incorporated in LE by
construction i.e., that is why the LE reflects it's properties. Since
coefficients $|a_{2M-i}|^2$ grow quadratically as $i$ approaches $M$
the most dominant cosine term in (\ref{LNLM1}) is for $i=N$ which
leads to the conclusion that the dominant frequency of revivals
$\omega_R$ for LE of $N<M$ particles is simply related to the SP
energy of $V_0(x)+V_l(x)$ potential through the relation
\begin{equation}
\omega_R(N)=\frac{E_{2M-N}^l-E_N^l}{\hbar}.\label{omega}
\end{equation}
Now consider the case $N=M$. Going from $N=M-1$ to $N=M$ particles, the
matrix (\ref{matrixP}) remains diagonal due to relation
(\ref{psiM}), we simply add $A_{MM}(t)\approx\exp(-iE_M^lt/\hbar)$
on the main diagonal, and since $|A_{MM}(t)|^2\approx1$, we get that the 
LE for $N=M$ is the same as the LE for $N=M-1$ particles, in accordance
with Fig. \ref{fig8} (red solid line for $N=40$ and red circles for
$N=39$ particles).

\begin{figure}[!h]\centering
\includegraphics[scale=0.40]{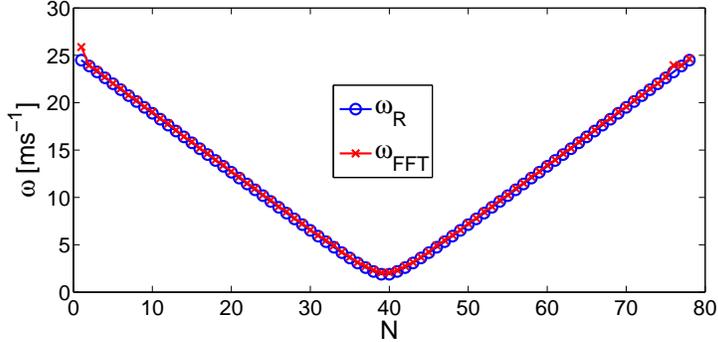}
\caption{Dominant revival frequency of Loschmidt echo obtained with
Eqs. (\ref{omega}) and (\ref{omega2}) $\omega_R$ (blue circles) and
with the Fourier transform of Loschmidt echo $\omega_{FFT}$ (red
crosses). Parameters used for calculation of LE are the same as in
Fig. \ref{fig8}. See text for details.}\label{fig9}
\end{figure}

Now we proceed to the $N=M+1$ case. In this case the matrix (\ref{matrixP})
obtains two off diagonal elements
\begin{equation}
A_{M+1,M-1}(t)=A_{M-1,M+1}(t)\approx\frac{-a_{M+1}\exp(-iE_{M+1}^lt/\hbar)-a_{M-1}\exp(-iE_{M-1}^lt/\hbar)}{1+|a_{M+1}|^2},\label{prop1}
\end{equation}
where we used (\ref{psi}) and $|a_{M-1}|=|a_{M+1}|$, see
(\ref{coeff}). The determinant of the matrix (\ref{matrixP}) has two
terms, the product of diagonal elements and a term arising from two
off diagonal elements
$$\det{\bf A}(t)\approx\prod_{i=1}^{M-2}A_{ii}(t)A_{MM}(t)\left[A_{M-1,M-1}(t)A_{M+1,M+1}(t)-A_{M+1,M-1}(t)A_{M-1,M+1}(t)\right].$$
It can be shown that
$$\left[A_{M-1,M-1}(t)A_{M+1,M+1}(t)-A_{M+1,M-1}(t)A_{M-1,M+1}(t)\right]\approx\exp\left[-i(E_{M-1}^l+E_{M+1}^l)t/\hbar\right],$$
which together with $A_{MM}(t)\approx\exp(-iE_{M}^lt/\hbar)$, yields
$$L(t)=|\det{\bf A}(t)|^2\approx\prod_{i=1}^{M-2}|A_{ii}(t)|^2.$$
We conclude that the Loschmidt echoes for $N=M+1$ and $N=M-2$ are
approximately the same. We see that the same pattern emerges as with
ground state fidelity. The Loschmidt echo values come in pairs, i.e,
it is approximately the same for $N=M-j$ and $N=M+j-1$ particles,
where $j=1,\ldots,M-1$. Due to this pattern, we can use Eq.~(\ref{omega}) 
to get the dominant revival frequency of Loschmidt
echo for other particle numbers $N\geq M$, i.e we can write
\begin{equation}
\omega_R(M+j-1)\approx\omega_R(M-j),\label{omega2}
\end{equation}
where $j=1,\ldots,M-1$ and $\omega_R(M-j)$ on the right side of Eq.
(\ref{omega2}) is given by Eq. (\ref{omega}).
In order to check the quality of these relations, we plot in Fig.~\ref{fig9} 
the dominant revival frequency $\omega_R(N)$ obtained via
(\ref{omega}) and (\ref{omega2}) for $N=1,\ldots,2M-2$ particles
(for parameters used here $M=40$), together with the most dominant
frequency $\omega_{FFT}(N)$ obtained with the Fourier transform of
LE; the agreement is excellent. Parameters used to calculate the LE are
the same as in Fig. \ref{fig8}.

\newpage
\subsection{Harmonic oscillator}

In this subsection we use Eqs. (\ref{LETG}) and (\ref{matrixP}) to
explore the LE of a TG gas following a sudden quench with optical
lattice $V_l(x)$; before the quench the TG gas is in the ground
state $|\Psi_0\rangle$ of harmonic oscillator potential $V_0(x)$
(\ref{HO}). In this subsection we use same units and parameters as
in Section \ref{sec:HO}.

\begin{figure}[!h] \centering
\includegraphics[scale=0.4]{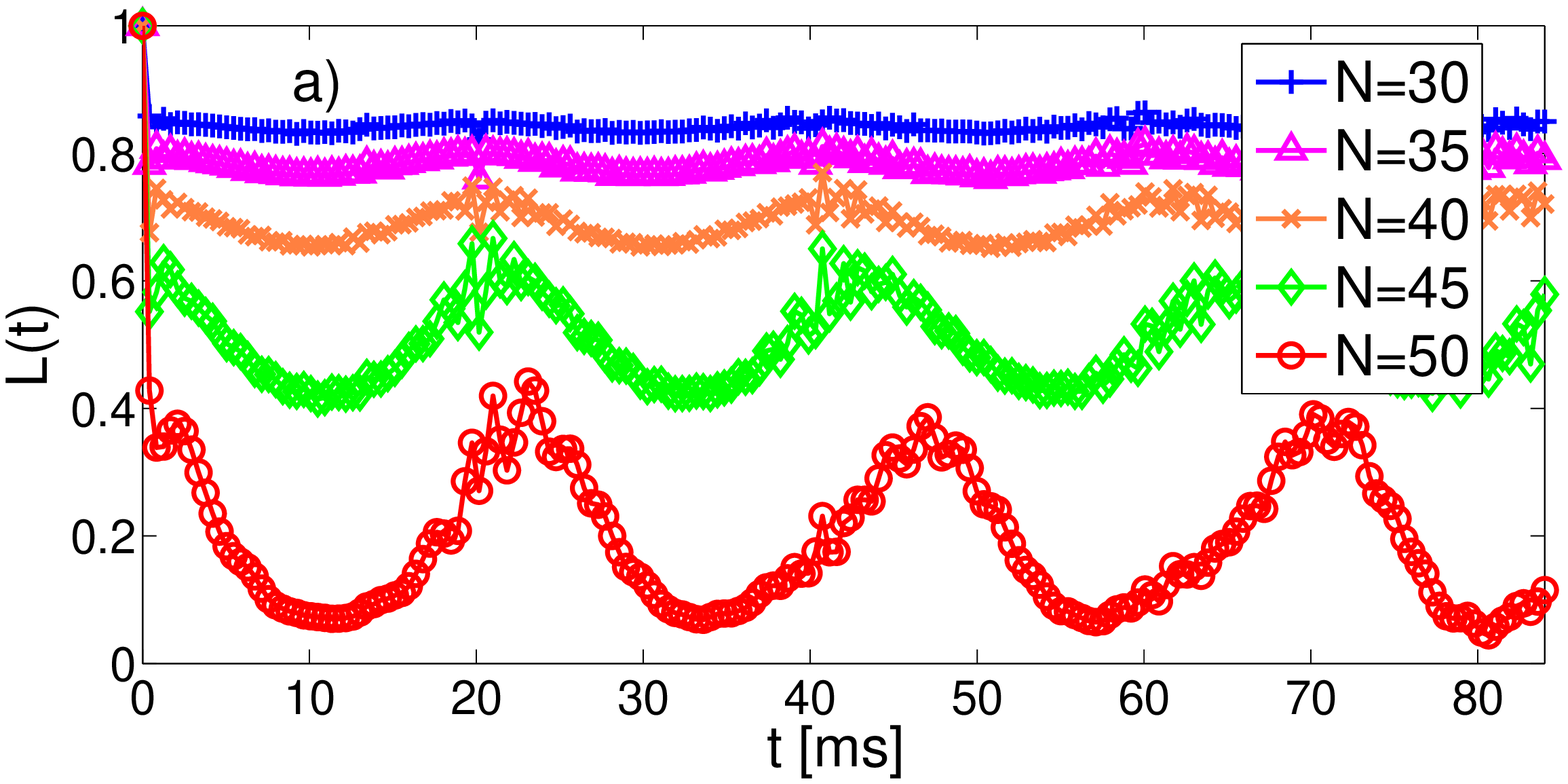}
\includegraphics[scale=0.4]{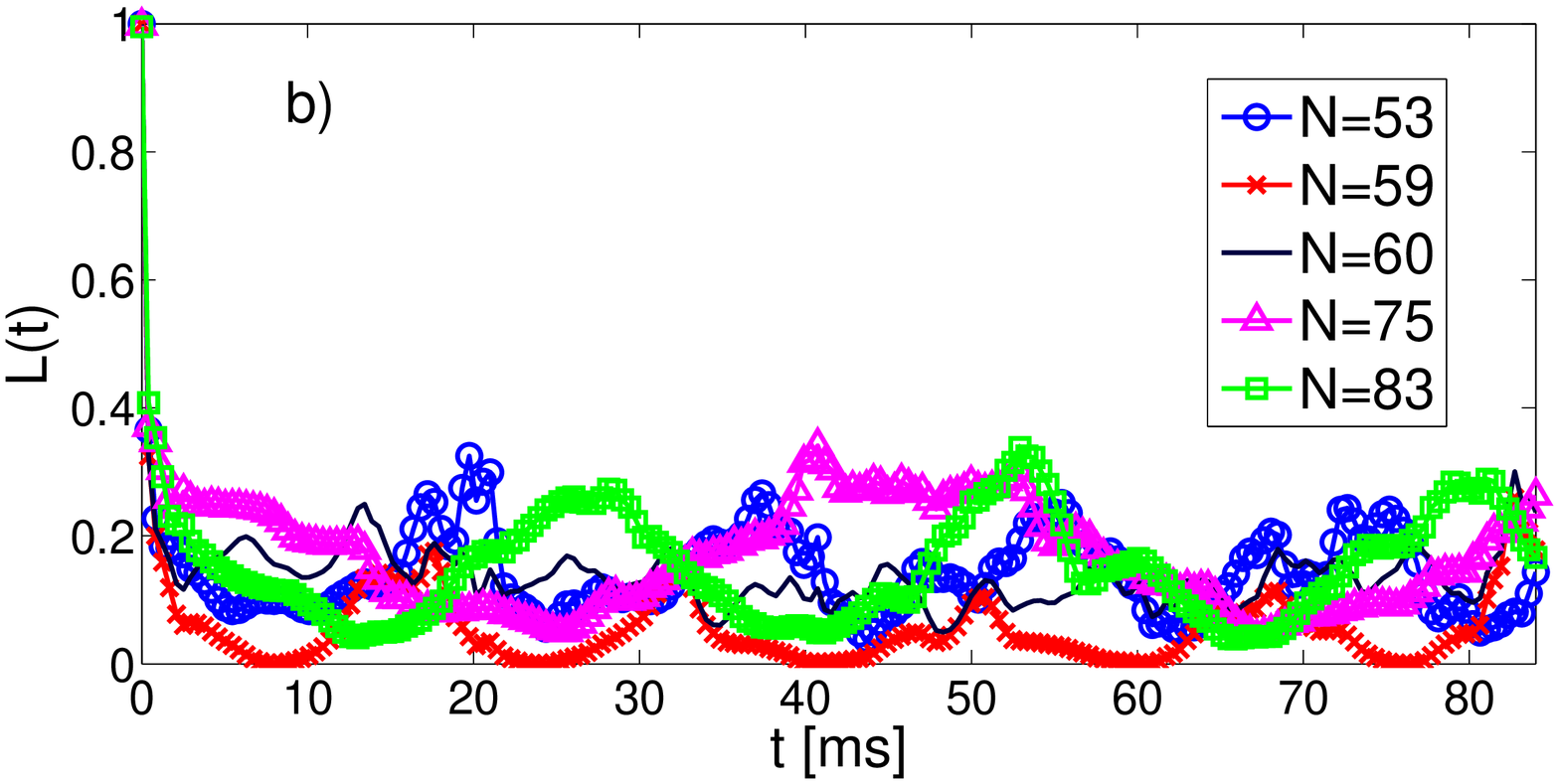}
\caption{The Loschmidt echo $L(t)$ for different particle numbers
$N$. a) LE for particle numbers up to $N<N_{pinn}$. b) LE for
$N>N_{pinn}$. See text for details.}\label{fig10}
\end{figure}

In Fig.~\ref{fig10} we show the Loschmidt echo following the quench
with optical lattice $V_l(x)=0.45E_R\sin^2(1.88\pi x_0^{-1}x)$; the
system was initially in a ground state of the harmonic oscillator
potential with $\omega_0/2\pi=25$Hz. Fig. \ref{fig10}(a) is for
$N<N_{pinn}$ and Fig. \ref{fig10}(b) is for $N>N_{pinn}$, where
$N_{pinn}$ is given by Eq.~(\ref{NHO}) (for parameters used here
$N_{pinn}\approx52$). We see that properties of GSF (see Fig.
\ref{fig6}(a) blue crosses) are reflected in the LE; the average
values of the LE are lower for lower GSF. This is a general
observation. However, the details of LE dynamics (such as the
dominant revival frequency) depend on trapping potential.

\begin{figure}[!h] \centering
\includegraphics[scale=0.4]{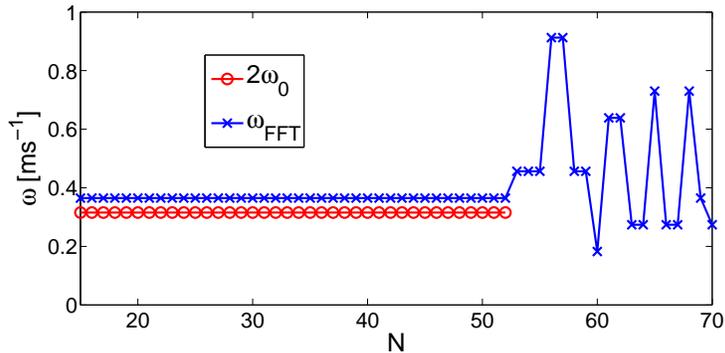}
\caption{ Dominant revival frequency of Loschmidt echo obtained with
Eq.~(\ref{omegaHO}) $\omega_R$ (red circles) and with the Fourier
transform of the Loschmidt echo $\omega_{FFT}$ (blue crosses).
The parameters used to calculate the LE are the same as in Fig. \ref{fig10}.
See text for details.}\label{fig11}
\end{figure}

In Fig. \ref{fig11} we illustrate the dominant frequency of the LE
(from Fig. \ref{fig10}) obtained with the Fourier transform
($\omega_{FFT}$ blue crosses). We see that $\omega_{FFT}$ is
constant for $N\leq N_{pinn}$, and it starts to behave irregularly
for $N>N_{pinn}$. We have found numerically that the regular
behavior for $N\leq N_{pinn}$ occurs because $\psi_n(x,0)$ can be
well approximated with
$$\psi_n(x,0)\sim A\phi_{n-2}(x)+B\phi_n(x)+C\phi_{n+2}(x),$$
where coefficient $B$ is always the largest in magnitude; this
yields
\begin{equation}
\omega_R(N\leq N_{pinn})\approx2\omega_0\label{omegaHO}
\end{equation}
which is also plotted in Fig. \ref{fig11}. For $N>N_{pinn}$ there
are many coefficients in the expansion of $\psi_n(x,0)$ in terms of
$\phi_n(x)$ which contribute on equal footing and simple relation
(\ref{omegaHO}) does not hold anymore.


\section{Conclusion}

We have studied the pinning quantum phase transition in a
Tonks-Girardeau gas, both in equilibrium and out-of-equilibrium,
using the ground state fidelity and the Loschmidt echo as diagnostic
tools. We have found, both numerically and analytically (within
first order perturbation theory), that the ground state fidelity can
individuate the region of criticality. The ground state fidelity defined
in Eq.~(\ref{FTG}) has a dramatic decrease when the atomic density
approaches the commensurate density of one particle per lattice
well. This decrease is a signature of the pinning transition from
the Tonks to the Mott insulating phase. We have found that the ground
state fidelity of the TG gas in an infinitely deep well potential can
be insensitive to finite size effects. The GSF for $N=M-1$ and $N=M$
particles ($M$ denotes number of lattice wells) in cosine-squared
(sine-squared) lattice are the same, while the single particle energy
spectrum and density show that pinning actually happens for $N=M$
($N=M-1$, respectively). The GSF has an advantage over the density and
single particle energy spectrum in the thermodynamic limit, where it
dramatically shows where the pinning takes place for an
infinitesimally small lattice amplitude. We have studied the
applicability of the fidelity for diagnosing the pinning transition
in experimentally realistic scenarios. Our results are in excellent
agreement with recent experimental work \cite{Nagerl:10}.

We have found that the GSF in a harmonic oscillator potentials shows
enhanced sensitivity in a broad region of particle numbers $N\geq
N_{pinn}$ (where $N_{pinn}$ is defined in Eq.~(\ref{NHO})); at
$N\sim N_{pinn}$ GSF has a faster decay and for larger $N$ develops
oscillations. This behavior is related to series of 'gaps' opening
at $n\sim N_{pinn}$ in the single particle energy spectrum of the total
potential (harmonic oscillator plus optical lattice).

In addition, we have explored the out of equilibrium dynamics of
the gas following a sudden quench with a lattice potential
potential. We have showed that all properties of the ground state
fidelity are reflected in the Loschmidt echo dynamics i.e., in the
non equilibrium dynamics of the Tonks-Girardeau gas initiated by
sudden quench of the lattice potential. The average value of the Loschmidt
echo is lower for lower values of ground state fidelity, regardless
of the details of the trapping potential. Details of the Loschmidt echo
dynamics, such as dominant revival frequency, depends on the type of
trapping potential. We find regular behavior of revivals for all
relevant particle numbers in infinitely deep well potential i.e,
frequency's get lower as we approach criticality and can be
calculated simply from single particle energy spectrum of the total
potential (infinitely deep well plus optical lattice). In the
harmonic oscillator potential, the dominant frequency of revivals
behaves in a regular way. It is a constant, approximately equal to $2\omega_0$
($\omega_0$ is frequency of harmonic trap), until a series of 'gaps'
open in the single particle energy spectrum of the total potential.

\acknowledgments

This work is supported by the Croatian Ministry of Science (Grant
No. 119-0000000-1015). H.B. acknowledge support from the
Croatian-Israeli project cooperation and the Croatian National
Foundation for Science. K.L and T.\v S are grateful to Ivana
Vuksanovi\'c for useful discussions. JG would like to acknowledge funding from an IRCSET
Marie Curie International Mobility fellowship.

\end{document}